# SIMULATIONS OF GALAXY FORMATION IN HIERARCHICALLY CLUSTERING UNIVERSES

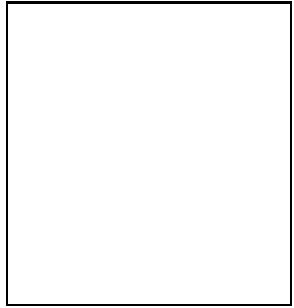

JULIO F. NAVARRO

*Steward Observatory, University of Arizona, Tucson, AZ 85721*

**Abstract**

I review the results of recent N–body/hydrodynamical simulations of the formation of galaxies in hierarchical universes, with special emphasis on the structure of dark matter halos, as well as on the mass and angular momentum of gaseous disks that assemble at the center of these halos. The density profiles of all dark matter halos can be fit by scaling a simple "universal" profile, regardless of the mass of the halo, the spectrum of density perturbations or the value of $\Omega$. Matching the observed rotation curves of disk galaxies with this halo structure requires disk mass-to-light ratios to increase systematically with luminosity. It also suggests that the halos of bright galaxies depend only weakly on galaxy luminosity and have circular velocities significantly lower than the disk rotation speed. This explains why luminosity and dynamics are uncorrelated in observed samples of binary galaxies and of satellite/spiral systems. For galaxy clusters, these halo models are consistent both with the presence of giant arcs and with the observed structure of the intracluster medium, and they suggest a simple explanation for the disparate estimates of cluster core radii found by previous authors. Simulations that include a radiatively cooling gas component show that the gas follows the dark matter evolution at early times, and that most of it collects at the bottom of the potential wells of the first resolved halos soon after they collapse. There, it settles into tightly bound cores that are hardly disrupted by subsequent evolution. As their surrounding halos coalesce, these gaseous cores merge on a time scale that is consistent with estimates based on dynamical friction considerations. This simple galaxy formation model suffers from serious shortcomings. It predicts that most of the baryons of the universe should be locked up in galaxies, in disagreement with observations. In addition, the spin of the simulated galaxies is well below that inferred from observations of spiral galaxies because of large angular momentum losses incurred during the merger events that characterize hierarchical galaxy formation. Only if feedback effects from evolving stars and supernovae dominate the process of assembly of galaxy disks can the observed dynamics of disk galaxies be reconciled with the hierarchical buildup of the dark halos that surround them.

## 1 Introduction

It is commonly assumed that structure in the universe grows hierarchically by gravitational instability of small density perturbations present at early times; small scale structures collapse first and then merge into more massive systems in a highly non-linear process best followed with numerical simulations. Once the spectrum of density fluctuations and the cosmological parameters have been specified, the evolution of the collisionless dark matter component can be followed in detail using standard $N$-body techniques. Explicit inclusion of the observable component of the universe (gas and stars) involves solving the equations which govern the evolution of a collisional fluid including the many additional physical processes which influence its structure: star formation and evolution; energy input from stellar winds, stellar radiation, and supernova explosions; ionization effects from stellar and QSO radiation fields; heavy element enrichment; heat conduction; and magnetic fields. Limitations in hardware and software, as well as in our understanding of the physics of these processes, have prevented us so far from attempting simulations that take into account all of these complex mechanisms. First attempts have considered only a select few, such as radiative cooling and photoionization effects, but the list keeps growing as faster computers become available and more sophisticated software is developed. Although incomplete, this approach has proved useful as a first step towards understanding the basic physics of the galaxy formation process. For example, high-resolution N–body simulations have led to a better understanding of the structure and formation of dark matter halos, while gasdynamical simulations that follow the evolution of radiatively cooling gas within an evolving population of dark halos have highlighted a number of successes as well as some potentially serious problems for hierarchical models of galaxy formation. In §2 I relate the results of a large series of N–body simulations intended to analyse the structure of dark matter halos formed in hierarchically clustering scenarios. In §3 I report on N–body/gasdynamical simulations of the formation of galactic disks. Section 4 summarizes the main conclusions of these studies.

## 2  The Structure of Dark Matter Halos

The density profiles of dark matter halos have long been thought to contain useful information regarding the cosmological parameters of the Universe and the spectrum of primordial density fluctuations. The pioneering analytic work of Gunn & Gott[18], for example, first suggested a cosmological significance for the observation that the rotation curves of disk galaxies are flat. In the early 80's cosmological N-body simulations confirmed and extended this analytic work, suggesting that values of the density parameter $\Omega$ close to unity and a density perturbation spectrum such as that expected if the universe were dominated by Cold Dark Matter were favoured by the "isothermal" structure of dark matter halos implied by flat rotation curves[15],[16],[33]. Since then dark matter halos have often been modeled as non-singular isothermal potential wells whose depth is usually identified with the velocity dispersion of stars in spheroidal galaxies or with the disk rotation speed in spirals. Since velocity dispersion increases with galaxy luminosity, this implies that more massive halos should surround brighter galaxies, a hypothesis which does not seem to be supported by observations. In fact, studies of the dynamics of binary galaxies and of spiral/satellite samples have revealed a distinct lack of correlation between the dynamics and the luminosity of the system[38],[42]. A further problem with non-singular isothermal halo models arises from observations of gravitational arcs in galaxy clusters and of the X-ray emitting intracluster medium. X-ray cluster cooling flow models require large values of the core radius ($\sim$ 100-200 kpc[11]) whereas gravitational arc models favour much smaller values ($\sim 30-60$ kpc[17]). (I assume $H_0 = 50$ km s$^{-1}$ Mpc$^{-1}$ for all physical quantities quoted in this paper.) I explain below how these discrepancies can be reconciled if halos are modeled using a simple density law suggested by high-resolution N–body simulations.

### 2.1  Density Profiles of Cold Dark Matter Halos

A total of 19 halos with masses ranging from that of a rich galaxy cluster ($\sim 10^{15} M_\odot$) to that of dwarf galaxies ($\sim 10^{11} M_\odot$) were selected from large cosmological simulations of a standard biased ($b = 1.6$) CDM cosmogony and resimulated individually with higher resolution using a treecode. The numerical parameters were carefully chosen so that the numerical resolution of each simulated halo was the same, enabling meaningful comparisons between systems of very different masses. In particular, all halos at $z = 0$ have between 5000 and 10000 particles within the virial radius, $r_{200}$, (the radius where the mean density contrast is 200), and the gravitational softening is chosen to be $\sim 1\%$ of the virial radius in all cases.

Figure 1: (**a**–left) The typical density profiles of CDM halos. The leftmost (rightmost) system has a mass $M_{200} = 3 \times 10^{11} M_\odot$ ($M_{200} = 3 \times 10^{15} M_\odot$). Arrows indicate the value of the gravitational softening in each simulation. (**b**–right) The characteristic overdensity of halos as a function of mass. Masses are expressed in units of the current non linear mass corresponding to the standard biased CDM spectrum, $M_* = 3.3 \times 10^{13} M_\odot$.

Figure 1a shows the density profiles of four halos spanning four orders of magnitude in mass. The solid curves are fits using a density profile of the form suggested by Navarro, Frenk & White[29],

$$\frac{\rho(r)}{\rho_{crit}} = \frac{\delta_c}{(r/r_s)(1 + r/r_s)^2}. \tag{1}$$

(Here $\rho_{crit}$ is the critical density and $r_s$ is a "scale" radius.) Remarkably, all the profiles are very well fit by this simple functional form. If radii are expressed in units of the virial radius, $r_{200}$, (which determines the total mass of the system, $M_{200}$) there is a single free parameter in eq.(1); the characteristic overdensity $\delta_c$. Figure 1b shows that this overdensity correlates strongly with the mass of the halo; less massive systems tend to be denser than their larger counterparts. Such a trend is expected in hierarchical clustering scenarios such as CDM, since lower mass scales collapse at higher redshift and should therefore have typically higher characteristic densities. If we define the formation redshift of a halo of mass $M$ as the first time when half of its final mass was in progenitors with individual masses exceeding some fraction $f$ of $M$, we can compute analytically the expected dependence of $\delta_c$ on $M$ by assuming that the characteristic overdensity of a halo merely reflects the density of the universe at the time of formation, i.e. $\delta_c(M) \propto (1 + z_{form}(M))^3$. The curves in Figure 1b show that such identification provides a good description of the results of the numerical experiments for various values of the parameter $f$, lending support to the conclusion that the characteristic density of a halo is determined primarily by its formation redshift[30]. Remarkably, the same density profile (eq. 1) also seems to fit very well the structure of halos formed in different hierarchical clustering models, regardless of the mass of the halo, the spectrum of initial density fluctuations, and the value of the density parameter $\Omega$[31].

The circular velocity, $V_c(r) = (GM(r)/r)^{1/2}$, of all 19 halos is shown in Figure 2a as a function of radius. Consistent with eq.(1), $V_c$ rises near the center and declines near the virial radius. Over a wide range in radius $V_c$ is almost constant, in agreement with the results of previous, lower resolution studies[15],[16],[33]. Is this dependence of $V_c$ on radius consistent with the observed rotation curves of galaxy disks? To compute the rotation curve of disk galaxies with a given rotation speed forming within a CDM halo, only two parameters need to be specified; the disk stellar mass-to-light ratio, $(M/L)_{disk}$, and the halo mass or, equivalently, its circular velocity at the virial radius, $V_{200} = (GM_{200}/r_{200})^{1/2}$. (This is because, for a given rotation speed, observations constrain the luminosity and optical radius of a spiral disk, as well as the typical shape of the rotation curve[32].) Slight modifications to the halo structure caused by the presence of the disk can be taken into account by assuming that the halo responds adiabatically to the assembly of the galaxy[2],[4].

Figure 2: (**a**–left) The circular velocity as a function of radius for the 19 halos simulated in our series. The curves are truncated at the virial radius, $r_{200}$, of each system. (**b**–right) The result of matching the shape of observed disk rotation curves (dotted lines. The dashed lines are the halo contribution to the total circular velocity (solid lines) in each case. Radii are expressed in units of the optical radius of each galaxy, defined as 3.2 times the exponential radial scalelength. $(M/L)_{disk}$ values are in the $B$-band.

Figure 2b shows the values of these two parameters needed to match the typical rotation curves of spiral disks, as given in ref. [32]. Some trends are clear; faster rotating (brighter) disks require larger disk mass-to-light ratios, $(M/L)_{disk} \propto L^{1/2}$, and the asymptotic halo circular velocity is systematically lower than that of the disk. Furthermore, bright galaxies, i.e. those with $V_{rot} > 200$ km/s, are surrounded by halos of very similar circular velocity or, equivalently, of very similar mass. Indeed, disks with rotation speeds between 200 and 300 km/s should be surrounded by halos whose mean circular velocities differ by only $\sim 10\%$ and which can be up to a factor of two lower than the disk rotation speed. This result agrees well with the estimates of Zaritsky & White[42], who found from their study of the dynamics of satellite galaxies that the average circular velocity of the halos of bright spirals is between 180 and 200 km/s. The weak correlation between halo and disk circular velocity shown in Fig. 2b also provides a simple explanation for the lack of correlation found between luminosity and dynamics of binary galaxies and satellite/primary pairs. This result is especially encouraging, since the halo properties we infer were chosen to match the shape of the inner rotation curves, and do not use any information about dynamics at the much larger radii probed by binaries or satellite companions. The structure of dark halos surrounding spiral galaxies appears to be quite similar to that of Cold Dark Matter halos.

## 2.2 Core properties of galaxy clusters

I now examine the consequences of the halo structure discussed above for X-ray and gravitational lensing observations of galaxy clusters. An ongoing debate concerns the different estimates of the core radius obtained when an "isothermal" cluster potential is assumed in models of the X-ray emitting gas and of the giant gravitational arcs observed in many clusters.

The X-ray emitting intracluster medium is often approximated by the hydrostatic, isothermal $\beta$-model[5]. The density profile of the X-ray emitting gas is then of the form, $\rho_{ICM} \propto (1 + (r/r_c)^2)^{-3\beta/2}$. The core radius, $r_c$, is usually a sizeable fraction of the total extent of the emission, and $\beta$ is typically found to be $\sim 0.6$-$0.8$ [19]. The halo mass profile can then be derived by assuming that the gas is isothermal and in hydrostatic equilibrium. The halo then follows the same density law as the gas, with the same core radius and with outer slope parameter $\beta_{DM} = \beta/\beta_T$, where $\beta_T$ measures the ratio of the "temperatures" of the dark matter and of the gas, $\beta_T = \mu m_p \sigma_{DM}^2 / kT_{gas}$ ($\mu\, m_p$ is the mean molecular weight of the gas, $k$ is Boltzmann's constant, and $\sigma_{DM}$ is the halo velocity dispersion). This model is frequently used to interpret X-ray observations and to constrain the dark matter distribution near the center of clusters[19],[8]. Cooling flow models with a $\beta$-model mass profile have been successful at explaining a number of observations, including the central X-ray surface brightness "excess" and the drop in central temperature seen in systems with strong cooling flows. The core radius of the dark matter is typically inferred to be in the range $r_c \sim 100\text{-}200$ kpc[11].

Figure 3: The density profile of an isothermal gas (dotted line) in hydrostatic equilibrium within a dark matter halo whose structure is given by eq. 1 (solid line). The dashed line shows a fit using a $\beta$-model. The parameters $r_c = 0.1 \times r_{max}$ and $\beta = 0.7$ give an excellent $\beta$-model fit to the gas profile.

On the other hand, such large cores are ruled out by observations of giant arcs, which require core radii of order 20-60 kpc when a $\beta$-model is used to describe the lensing cluster[17]. This discrepancy is resolved if we assume instead that cluster halos follow the density profile of eq.(1). For the parameters shown in Figure 1b, halos are sufficiently concentrated to agree with the gravitational lensing constraints. A CDM cluster with mean velocity dispersion $\sim 1000$ km/s placed at $z = 0.3$ can produce giant arcs similar to those seen[36] yet it can also be consistent with a large core radius in the X-ray emitting gas. Requiring that the ICM be *isothermal* and in hydrostatic equilibrium in the dark matter potential (as suggested by numerical simulations[29]) results in a well defined core. This is illustrated in Figure 3, where we plot density profiles for an isothermal gas (dotted line) and for the dark matter (solid line). Radii are given in units of the radius where the circular velocity is maximum, $r_{max} \approx 2r_s$, and the density units are arbitrary. It is also assumed that the gas and dark matter temperatures are equal, $kT_{gas}/\mu\, m_p = \sigma_{DM}^2 = GM_{200}/2r_{200}$.

The difference in shape between the gas and the dark matter density profiles is due to our assumption that the gas is isothermal. Since the dark matter velocity dispersion drops at radii larger and smaller than $r_{max}$ (see Figure 2a), the gas structure deviates strongly from that of the dark matter at small and large radii. At small radii an isothermal gas develops a well defined core, and at large radii its density drops less rapidly than the dark matter. Such leveling off of the gas profile is not observed in real clusters, indicating that the gas temperature must decrease in the outer regions. Hydrodynamical simulations confirm that this is the case, and indicate that the gas and dark matter distributions in the outer regions are very similar[29]. The dashed line shows the result of fitting the gas profile with the $\beta$ model mentioned above. In this case, values of $r_{core} = 0.1 \times r_{max}$ and $\beta = 0.7$ give a very good fit to the structure of the gas. For a cluster with a velocity dispersion of 1000 km/s, this implies a gas core radius of about 120 kpc, in good agreement with observed core radii.

Detailed cooling flow models that use the potential corresponding to eq.(3) rather than the $\beta$-model also agree well with observation. A recent analysis by Waxman & Miralda-Escudé[36] shows that the observational signatures of cooling flows in CDM halos are essentially indistinguishable from those occurring in halos with a true constant density core. The apparent discrepancy between X-ray and gravitational lensing estimates of the core radius seems to be a direct result of force-fitting a $\beta$-model to systems whose structure is better described by a profile more similar to that of CDM halos.

Figure 4: Particle plots at various redshifts of a system with final circular velocity of about 120 km s$^{-1}$. The four panels on the left show dark matter particles. The right panels show the gas particles. A label in each plot gives the redshift. The region shown is 400 *physical* kpc in every case. Note that the gas collapses to the center of dark matter halos, where it settles into disk-like structures.

## 3   The Formation of Galactic Disks

The "standard" model of disk galaxy formation assumes that they form as a consequence of the dissipative collapse of baryons in the potential wells of hierarchically clustering dark matter halos, a process that can account naturally for the characteristic sizes, masses, and rotational properties of galaxies[34],[37],[13]. This scenario has, however, some serious shortcomings. Since dissipative effects are more efficient at high redshift (when all systems were in general denser and colder than today), it is necessary to postulate the existence of some mechanism that prevents most of the baryons from cooling and transforming into stars at early times[37],[6],[39]. A related problem is that the mass function of galaxies is predicted to have a much steeper slope at the low-mass end than the observed galaxy luminosity function. This disagreement with observations applies to all hierarchical clustering models, and can only be ameliorated if the formation of stars is severely suppressed in halos of small mass[22],[7]. These problems were identified adopting very schematic and simplified descriptions of the evolution of baryons in dark halos, and they need to be confirmed by N-body/hydro simulations. A number of such simulations have been reported recently, and they seem to confirm the problems stated above. I shall concentrate here on a few outstanding questions that can be addressed directly with these numerical experiments. What fraction of the gas in a galactic halo can cool and settle into a central galaxy? How does this fraction depend on mass and formation epoch? How does the angular momentum of a gaseous disk relate to that of the surrounding dark halo? How frequent are mergers, what is their influence on galactic structure and what constraints do they put on cosmological parameters?

The numerical experiments that I will discuss are very similar to the ones reported in §2, but include a gaseous component which is evolved using the Smooth Particle Hydrodynamics technique. The runs include the effects of gravity, pressure gradients, hydrodynamical shocks and radiative cooling. They neglect the effects of star formation and their possible feedback into the interstellar medium. (Readers should consult ref. [26], [27], [28], and [29] for details.) The dissipative collapse of the baryonic component that characterize the formation of galaxies in this model leads naturally to the formation of gaseous, rotationally supported disks whose size is determined by the angular momentum content of the gas[26] (Figure 4). Much of the gas is collected at high redshift at the center of protogalactic halos, forming rotationally supported tightly bound "cores" surrounded by a hot tenuous atmosphere of pressure supported gas. As halos collide and merge, the cores sink to the center of the new halo, eventually merging into a single object in a time scale consistent with simple dynamical friction considerations. For typical merging protogalactic halos, dynamical friction timescales are relatively short, and therefore most of the gas ($> 70\%$) is concentrated in the central core practically at all times.

The efficient cooling and early collapse of baryons to the center of dark halos seen in these simula-tions is very difficult to reconcile with the observation that galaxies contribute a very small fraction of the total baryonic density of the universe implied by nucleosynthesis constraints[6],[39]. This point is strengthened by the observation that galaxies contain only a small fraction of the total baryonic mass of rich galaxy clusters[40]. For a hierarchical model to be successful, it must contain some mechanism that prevents the gas from collecting at the center of low mass halos at high redshift, when radiative cooling times were extremely short. Mechanisms that could help solve this problem have been proposed, mainly based on the effect that feedback from supernovae and evolving stars may have on the interstellar medium[22],[7], or on the reduction of the efficiency of cooling caused by a photoionizing background[9].

Figure 5: (**a**–left) The specific angular momentum of halos and gaseous disks as a function of mass. The boxes show the locii of observed spiral disks and elliptical galaxies (Fall 1983). Note that gaseous disks have angular momenta which are too small to compare favourably with spiral disks. (**b**–right) The fraction of systems that have accreted more than 5 and 10 percent of their mass over the lookback time indicated. The solid and dotted lines refer to the dark matter halos and gaseous disks, respectively. Note that disks accrete mass much less efficiently than do their surrounding halos.

A related problem that arises if baryons are collected at the cores of protogalactic halos at early times and accrete most of their mass through mergers is that substantial losses of angular momentum accompany the mergers of these cores[15],[43],[1]. Figure 5a shows the specific angular momentum of halos (filled circles) and gaseous cores (open circles) as a function of mass. The specific angular momentum of gaseous disks is, on average, only a tenth of that of their surrounding halos, although tidal torques gave both components approximately the same spin during the expansion phase. (The halo and disk particles had approximately the same $J/M$ at maximum expansion.) As a result, typical spiral galaxies have higher angular momenta than the gaseous disks formed in these simulations, a fact already noticed by Navarro and Benz[24] and Navarro and White[26]. It is possible that the same mechanism that prevents the baryons from condensing at the center of small halos at high redshift can change the dynamics of the collapse of the gas, allowing it to conserve a larger fraction of its initial angular momentum.

The concentration of the gaseous cores relative to their surrounding halos can also cause their subsequent dynamical evolutions to decouple. This dynamical decoupling between galaxies and halos may have a profound influence on the timing and mode of galaxy formation. For instance, *galaxy* mergers may be delayed relative to the merger of halos. (Indeed, this is a requisite for the viability of hierarchical models since it can prevent the formation of supermassive galaxies in halos with masses corresponding to galaxy groups and clusters[37],[16],[21],[20],[10].) This is particularly important because many of the observed properties of a galaxy are likely to depend quite sensitively on the time and manner in which its mass was assembled. For example, thin stellar disks are extremely fragile to the accretion of satellites. Tóth and Ostriker[35] estimated that a galaxy like our own must have accreted less than 10 percent of its disk mass in the past 5 Gyrs, a very low accretion rate apparently in conflict with the high *halo* merger rates expected in a universe with $\Omega = 1$. These authors interpret this as evidence against a high-density universe. However, dense satellite *cores* can take longer to merge with the central disk than their surrounding halos, an effect that can reconcile the large fraction of thin spirals seen today with large values of $\Omega$[27].

This can be seen in Figure 5b, where we show how much material has accreted recently onto halos and onto disks. Over the past 5 Gyrs (ie. since $z = 0.38$) about 70% of the dark halos have accreted at least 10% of their mass and more than 90% have accreted at least 5% (solid lines). The dotted lines correspond, on the other hand, to the fraction of disks that have grown by more than 5 and 10%. The difference between the solid and dotted lines indicates that disks are less efficient at accreting mass than are the dark halos that surround them. In fact, fewer than 30% of the disks grow by more than 10% in the past 5 Gyr, and much of the recently accreted gas is in the form of small compact satellites which are still in orbit at $z = 0$. Therefore, the fraction of galaxies that remain relatively undisturbed in our models over the last 5 Gyrs is comparable to the observed spiral fraction among field galaxies. Since the distribution of disk thickness for real galaxies is poorly known, a more detailed quantitative comparison is not possible at present. I conclude that dismissal of $\Omega = 1$ on the basis of the existence of thin stellar disks is perhaps premature.

A final problem concerns the rapid collection of baryons at the center of dark halos. As shown by Navarro and White[26], and confirmed by these simulations, mergers between gaseous cores proceed rather quickly *on galactic scales* and consequently most of the baryonic mass ($> 70\%$) is concentrated in the central core practically at all times. This is actually consistent with observations of individual galaxies, including our own Milky Way, where most of the baryons within the virial radius appear to be associated with the central galaxy. If this result is extrapolated to rich galaxy clusters, one would expect central cluster galaxies to contain a large fraction of the stars in the cluster. However, observations indicate that, even for cD clusters, the central galaxies contain typically not more than about 10-20% of the stellar mass of the cluster[3]. If the formation of clusters is just a "scaled-up" version of that of galaxies, what makes clusters different from galaxies in this respect? Rich clusters, the largest virialized systems present today, differ from galaxies in that they have been assembled only recently through the amalgamation of units which on average represent a very small fraction of the total mass. Since dynamical friction timescales are controlled by the *ratio* of satellite to primary halo mass, it is possible that this "bias", together with the very recent formation times of clusters, may be the reason why the merging process is today closer to completion in individual galaxies than in galaxy clusters. This is in qualitative agreement with the results of Kauffmann etal[22], and with the large numerical simulations of Katz etal[21] and Evrard etal[10], but still needs to be confirmed by numerical simulations of higher resolution than have been possible so far.

## 4   Conclusions

The results presented in the previous sections offer an interesting insight into how the mass of galaxies is assembled in a hierarchical universe. They confirm predictions of previous theoretical works, and show ways in which analytical frameworks that have been developed to study the evolution of dark halos can be extended in order to describe the way in which galaxies form within these halos. This work also highlights a number of problems that hierarchical models face when compared with observations. These problems should not be taken as excluding hierarchical clustering models in general. Rather they reinforce the conclusion that a successful model will require a better understanding of star formation and of how processes such as supernova heating or photoionization by an ultraviolet background can affect the dynamics of protogalactic gas. It is unlikely that a definitive picture of galaxy formation will emerge until these complex physical processes are understood in detail.

**Acknowledgements.** I am grateful to my collaborators, Carlos Frenk, and Simon White for allowing me to discuss the results of our work here.